# EEG-based Investigation of the Impact of Classroom Design on Cognitive Performance of Students

Jesus G. Cruz-Garza[1], Michael Darfler[1], James D. Rounds[2], Elita Gao[1], Saleh Kalantari*[1]
[1] Cornell University, Department of Design and Environmental Analysis
[2] Cornell University, Department of Human Development

*Corresponding author: Saleh Kalantari <sk3268@cornell.edu>

**Abstract**
This study investigated the neural dynamics associated with short-term exposure to different virtual classroom designs with different window placement and room dimension. Participants engaged in five brief cognitive tasks in each design condition including the Stroop Test, the Digit Span Test, the Benton Test, a Visual Memory Test, and an Arithmetic Test. Performance on the cognitive tests and Electroencephalogram (EEG) data were analyzed by contrasting various classroom design conditions. The cognitive-test-performance results showed no significant differences related to the architectural design features studied. We computed frequency band-power and connectivity EEG features to identify neural patterns associated to environmental conditions. A leave-one-out machine-learning classification scheme was implemented to assess the robustness of the EEG features, with the classification accuracy evaluation of the trained model repeatedly performed against an unseen participant's data. The classification results located consistent differences in the EEG features across participants in the different classroom design conditions, with a predictive power (test-set accuracy: 51.5%-61.3%) that was significantly higher compared to a baseline classification learning outcome using scrambled data. These findings were most robust during the Visual Memory Test, and were not found during the Stroop Test and the Arithmetic Test. The most discriminative EEG features were observed in bilateral occipital, parietal, and frontal regions in the theta (4-8 Hz) and alpha (8-12 Hz) frequency bands. Connectivity analysis reinforced these findings by showing that there were changes in the transfer of information from centro-parietal to frontal electrodes in the different classroom conditions. While the implications of these findings for student learning are yet to be determined, this study provides rigorous evidence that brain activity features during cognitive tasks are affected by the design elements of window placement and room dimensions. The ongoing development of this EEG-based approach has the potential to strengthen evidence-based design through the use of solid neurophysiological evidence.

# 1. INTRODUCTION

*1.1. Overview and research objectives*

The indoor environmental quality of classrooms has been linked to students' physical and mental health and learning outcomes. This includes factors such as the room's lighting, color-scheme, maintenance level, noise level (Klatte et al., 2010), temperature (Mendell & Heath, 2005; Wargocki & Wyon, 2007), furniture quality and layout (Castellucci et al., 2017), air quality (Daisey et al., 2003), and the incorporation of connections to nature (Kidger et al., 2012; Ulrich, 2006). In addition to ambient environment variables, the design of the physical classroom, such as seating arrangement (Wannarka & Ruhl, 2008), seating type (Harvey & Kenyon, 2013), access to technology (Radcliffe, 2008), color (Engelbrecht, 2003), and building quality (Maxwell & Schechtman, 2012) have been shown to impact student attitudes, achievements, and behaviors, even when controlling for socioeconomic variables.

      However, this prior literature on student health, behaviors, and outcomes has not focused as extensively on specific measurements of cognitive performance in relation to the indoor environmental features of classrooms. Cognitive performance refers to the fluency, efficiency, and accuracy of carrying out cognitive tasks (Demetriou et al., 2020). Higher cognitive performance indicates more effective reasoning, problem-solving, and decision-making skills, and has been associated with better information recall and long-term mental acuity (Salthouse, 2005). Researchers have shown that some environmental features, such as suboptimal thermal conditions and poor air quality, can impair cognitive performance, most likely through an interconnected combination of physical pathways and psychological distraction (Adam et al., 2008; Bandelow et al., 2010; Mäkinen et al., 2006; Martin et al., 2019; Morley et al., 2012; Parker et al., 2013; Spitznagel et al., 2009; Taylor et al., 2014; Zhang et al., 2019; Zhu, Liu, & Wargocki, 2020). The data on these readily measured environmental quality factors is compelling, but much less is known about the ways in which specific architectural features, such as room size, window locations, or furniture layout, may impair or improve cognitive performance.

      The growing movement toward evidence-based approaches in architectural design has yielded promising results, but this is a challenging field in which to create scientifically robust studies (Cama, 2009; Edelstein & Macagno, 2012; Hamilton & Watkins, 2008). The current study's approach is commensurate with an incipient neurocognitive direction in the field, in which researchers seek not only to gather solid behavioral and performance data but also to describe the underlying brain mechanisms that mediate human responses to architectural features (Vartanian et al., 2013). In recent years several "neuroarchitectural" studies have investigated brain responses to different architectural styles (Choo et al., 2017), contours (Vartanian et al., 2013), height and enclosure (Vartanian et al., 2015), architectural façade and landmark recognition (Rounds et al., 2020), interior space design (Banaei et al., 2017), lighting (Shin et al., 2015; Lu et al., 2020), ambient conditions (Choi, et al., 2015; Guan et al., 2020; Choi, et al., 2019), and color (Küller et al., 2009). Other researchers have taken a neural approach to examine the impact of the built environment on human memory (Sternberg, 2010, p. 147) and comparative human responses to built vs. natural environments (Banaei et al., 2017; Roe et al., 2013). In the context of the learning environment, Kim et al. (2020) used EEG measurement to investigate the effect of indoor thermal conditions on college students' learning performance.

      Due to a lack of opportunities (and the very high expense) for enacting precise architectural changes in real-world structures, the most scientifically robust studies in this area have been carried out by measuring participants' neural responses while they view 2D images of architectural designs. As a result, the existing neurological research may not fully capture the relevant brain responses associated with experiencing an immersive three-dimensional environment (Coburn et al., 2017; Vecchiato et al., 2015). While virtual-reality (VR) experiences are also limited in their ability to mimic embodied engagements with architecture, they do allow for a more dynamic and immersive experience and a much

richer sense of presence, compared with 2D stimuli (Makransky et al., 2020). Research that uses an immersive, three-dimensional VR approach is thus likely to capture a more accurate reflection of brain responses to these environments compared to studying participants who are looking at two-dimensional pictures. The use of VR in the current study allowed the researchers to precisely adjust architectural variables in a way that would not be feasible in real-world contexts (Kalantari & Neo, 2020), for example by adding different window placements in an otherwise identical classroom. It also allowed us to hold constant other environmental variables that may strongly affect neural processes, such as temperature, noise, and air quality.

In a previous study from our group (Kalantari et al, 2021), we compared EEG data taken while participants completed cognitive tests in a real classroom vs. in an identical virtual classroom. The results indicated that EEG frequency band-power was not significantly altered between the real-world and virtual environments. Early results show a promising indication that today's high-resolution, immersive VR technologies can provide a good proxy for estimating neural responses to in-person, real-world architectural design factors (Kalantari et al., 2020; Kalantari et al., 2019; Kjellgren & Buhrkall, 2010; Palanica et al., 2019; Yin et al., 2018, 2020).

The current study is to the best of our knowledge the first that uses EEG data to evaluate neural responses to precise changes in window placement and room width. We have also adopted a comprehensive approach to analyzing this data to find robust and generalizable brain-activity features associated with performing cognitive tasks in rooms with different architectural design factors. We analyze a combination of the two common types of features used to represent EEG time signals (Lotte et al., 2018): frequency band-power features (Ravindran et al., 2019; Rounds et al., 2020), and connectivity between electrodes (Cruz-Garza, 2020) at the scalp level. The current research takes a data-driven, bottom-up approach to EEG analysis by using feature extraction algorithms and machine learning. This technique is a post-hoc analysis that seeks to identify the specific brain measurements that contain the most relevant and least redundant information (i.e., what neural signals have changed the most in the different classroom conditions). The machine-learning algorithm thus allows the researchers to find the most relevant neural data, i.e. those neural patterns most affected by specific architectural design changes, while minimizing the bias introduced by a priori assumptions.

## 1.2. Prior research on the impact of classroom windows and room dimension on cognitive performance

Received wisdom has long held that students gazing out of windows are likely to be neglecting their studies. Recent research has challenged this assumption, suggesting that when students of many different ages have access to window views in their classrooms they tend to exhibit both reduced stress and increased attention to educational materials (Bagot et al., 2015; Collado & Corraliza, 2015; Faber Taylor & Kuo, 2009; Li & Sullivan, 2016; Matsuoka, 2010). Empirical testing in this area remains limited, particularly due to the difficulties in isolating window placement variables apart from other confounding classroom factors. In addition to the presence or absence of a window, there is also the question of the effects of different window locations, such as in the front of the classroom, to the side of students, and/or above the line of sight. The variable of what kind of view can be seen from the window (natural scenery, a parking lot, a busy street, etc.) is also likely to be an important factor. In the current study three different window options were evaluated: (a) no window view, designated as the "neutral" condition, (b) a side view tangential to the instruction, and (c) direct forward views located behind the instruction area. In both of the window conditions (b and c), the same image of natural exterior scenery was used, consisting of a view of tree-tops and clouds with no motion distractions, buildings, or other human-made elements.

Views of nature were chosen for the current study because they have been previously associated with increased cognitive performance in dormitory, classroom, and workplace settings (Barrett et al., 2015; Benfield et al., 2015; Chen, 2016; Tennessen & Cimprich, 1995). Other researchers have found that views to nature can mitigate the performance-eroding effects of unpleasant environmental conditions such as hot temperatures (Ko et al., 2020). However, the window locations, exterior view conditions, and other environmental variables were not controlled in these prior studies. The predominant theoretical basis that helps explain the beneficial effect of natural views on cognitive performance is Attention Restoration Theory (Kaplan, 1995). This outlook holds that the ability to maintain mental focus can become fatigued or depleted through sustained use, and that this capacity can be restored through the relaxed attention promoted by natural environments (Kaplan, 1995, p. 70).

As with other architectural design variables, empirical research on how room dimensions affect student performance has proven difficult to quantify due to the unique and complex construction of different classrooms. Most of the work in this area is several decades old and is based on less than robust methodologies. For example, Moore et al. (1996) speculated that ceiling heights of less than eight feet would probably lead to more subdued and focused behavior among students, whereas ceilings above eight feet would probably lead to more active and social behavior. These conclusions were based on observations of child-care centers and interviews with caretakers. Read and colleagues (1999) found that lower ceiling heights were associated with greater cooperative behavior levels among preschool children—a conclusion that somewhat contradicts Moore and colleagues (1996). Many other researchers identified ceiling height as an important variable that is likely to affect student performance, but did not make any specific suggestions about what that relationship might be (Glen I. Earthman, 2004; Olds, 1989; Winchip, 1991). To the best of our knowledge, there is no research literature at all on room width dimensions and cognitive performance, though one early study found that smaller rooms were associated with greater stress, compared to larger rooms with the same number of occupants (Sundstrom, 1975).

While adjusting classroom widths and the ceiling heights has been suggested by various researchers as a way to add value to the learning environment, it is difficult to ascertain, on the basis of the existing evidence, in which direction it should be altered, what specific effects could be expected from dimensional changes, and what confounding variables such as crowdedness, empty space, or symmetry might contribute. In the current study, two options for the classroom dimensions were evaluated: (a) narrow classroom of 15 ft. by 50 ft., designated as the "neutral" condition, and (b) wide classroom of 22 ft. by 50 ft. The ceiling's height was not manipulated in the current study; it was held constant at 12 ft. in each of the classroom variants. The room size (width, length, and height) for neutral condition was selected based on the most common classroom size in the site of the current study.

*1.3. List of hypotheses*

Hypothesis 1. Isolated changes in architectural design variables in a virtual classroom setting will be associated with significant effects in EEG signatures, as well as with cognitive task performance.

Hypothesis 2. Direct window views to natural scenery added behind the instruction area will improve cognitive task performance and will have a significant effect on EEG features.

Hypothesis 3. Views to natural scenery from a window added on the side-wall will improve cognitive task performance and will have a significant effect on EEG features.

Hypothesis 4. Expansion of the classroom's width will improve cognitive task performance and will have a significant effect on EEG features.

## 2. MATERIALS AND METHODS

All participants gave informed written consent before participating in the experiment and were compensated with a $25 gift card at the end of the study. The overall study protocol was approved by the Institutional Review Board at the University of Houston.

*2.1. Participants*

Twenty-nine healthy human adults with normal or corrected-to-normal vision were recruited for the study using a convenience sampling method (word-of-mouth and announcements on departmental e-mail lists).

Five of the participants had to be excluded due to technical issues during the physiological data-recording (missing/incomplete data), and two additional participants were excluded due to extremely low task accuracy scores (< 10%) across all conditions. The remaining 23 participants ranged in age from 18 to 55 years (M = 26.17, SD=10.4). The majority were university students (n=21), the other participants were faculty members. Eight participants identified as female, and 13 identified as male, and none as other. All of the participants were associated with the University of Houston, representing departments in Architecture, Engineering, Biological Sciences, Humanities, Economics, and Computer Sciences. The participants had a variety of national backgrounds—U.S., India, and Mexico were the most common—as well as different ethnic backgrounds. Eleven participants reported as Asian, seven as Latinx and/or Hispanic, one as Middle Eastern, and four as White non-Hispanic.

The participants were also asked to report their sleep patterns and rate their mental fatigue level at the beginning of the experiment. Twenty-six percent reported getting seven or more hours of sleep the previous night, 30% reported between six and seven hours of sleep, and 44% reported four to six hours of sleep. Reported fatigue levels were relatively low (M=4.1, SD=1.9, on scale from 1 = low fatigue to 10 = high fatigue). Participants also verified that to the best of their knowledge they were free of current neurological conditions, and had abstained from any psychoactive substances prior to the study (except for two who had recently consumed caffeine, within 1 hour prior to the study).

Participants were asked in a post-experiment survey to indicate their level of stress or discomfort with the virtual reality experience and the EEG headset, (1=comfortable with no stress; 10 = uncomfortable and stressful), and to report how realistic the virtual environment seemed (1=not similar to reality; 10=very similar to reality). Reported discomfort levels were relatively low for the VR system (M=4.1, SD=2.3) and for the EEG headset (M=3.8; SD=2.3). Participants found the VR environment to be reasonably realistic (M=6.2; SD=1.5).

## 2.2. VR development

The virtual environment was developed using Epic Games' Unreal Engine (www.epicgames.com), which is one of the most sophisticated virtual-reality simulation programs available today. The Unreal Engine uses blueprint scripting and thus allows for a quick learning curve on the part of researchers and designers, making the construction and modification of the new virtual-reality architectural models a relatively easy process. For each of the different classroom conditions, one architectural element of the control environment was modified (the window placement or room width). A widget was used to display cognitive tests on a virtual projector screen at the front of the classrooms (**Figure 1**). The experience was presented to the research participants using an Oculus Rift head-mounted display, a lightweight, high fidelity headset with a strong market share. The headset uses a 1920 x 1080 pixel (960 x 1080 pixel per eye) low-persistence OLED display with a refresh rate up to 75 Hz.

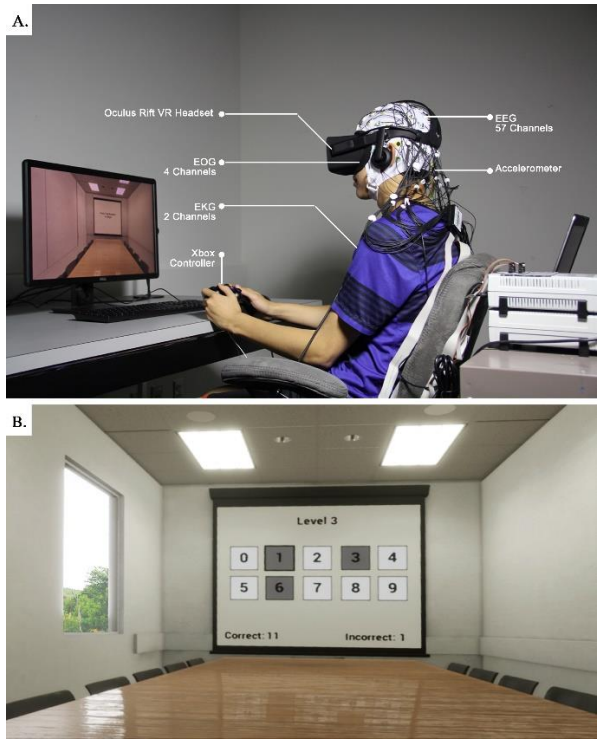

**Figure 1. A.** The virtual-reality headset along with EEG, EOG, and EKG as worn by a study participant; and **B.** a two-dimensional capture from the actual immersive 3D experience, showing the Digit Span Test in the virtual classroom with a side-window.

### 2.3. Classroom design conditions

Four design conditions were evaluated. First, the researchers constructed an exact VR rendering of a real university classroom with no windows; this was designated as the "neutral" or default classroom. Second, the neutral classroom was altered by adding one window to the left-hand wall, with a view of partly sunny skies and treetops. Third, the neutral classroom was altered by extending the room width and adding two windows at the front of the room, with a similar tree-top view (no side-window in this condition). Fourth, the neutral classroom was altered by extending the front wall to create a wider space, with no added windows (**Figure 2A**). All four of these design conditions were experienced by each participant; the order in which they appeared for any given participant was randomized.

### 2.4. Cognitive tests

In each of the four classroom design conditions, the participants performed five cognitive tests (**Figure 2B**). The selection of the tests was based on neuropsychological tests commonly used in literature related to cognitive performance evaluation among students (e.g. Nigg et al., 1999; Kim et al., 2020). These were presented in the same order in each classroom variation: the Digit Span Test (DST), the Benton Test (BT), the Stroop Color Interference Test (ST), an Arithmetic Test (AT), and then a Visual Memory Test (VMT) (Kalantari et al., 2020). The DST, BT, and VMT tests ended when a certain number of responses were submitted (22, 4, and 22, respectively). The ST and AT were time-limited at 25 s.

The DST (Wechsler, 1981) is an instrument frequently used by researchers to assess working memory, attention, encoding, and auditory processing. Participants listened to a spoken sequence of random digits and then repeated the sequence by selecting the correct digits on the screen. The study used four sequences of progressively greater lengths (four, five, six, and seven digits). Participants were given visual and auditory feedback indicating correct and incorrect answers. There was no time limit for this test. Participants were scored based on the total number of correct digits (out of 22 total possible through all four sequences) and on their longest complete span of correct numbers (up to a maximum of seven).

The BT (Benton, 1945; Sivan, 1992) is a measure of visual perception, visual memory, and the capacity for visual discrimination. Participants were shown a complex geometric visual stimulus for three seconds, followed by a set of four visual options (correct answer plus three similar variations), and were asked to select the one that was identical to the original stimulus. Each BT session for a classroom condition included four trials, each showing a different geometric test stimulus, and different incorrect options on the response screen. No feedback was given about the correct or incorrect answers. There was no time limit. Participants were scored on the number of correct trials (out of 4).

The ST consists of quickly identifying the font color of a textually congruent or incongruent word presented on a screen; for example the word "red" presented in a green-colored font represents the incongruent condition (Stroop, 1935). Words were presented serially, with a new word presented immediately after each response; the frequency of congruent and incongruent conditions was balanced, and the words and text colors were counterbalanced in order across sessions. Participants were given 25 seconds and asked to complete as many trials as possible. Visual and auditory feedback was given for correct and incorrect answers. Participants were scored on the ratio of correct answers to total attempted answers.

The AT was developed by the current researchers for this study based on a commonly used quantitative skills format; it presented participants with two simple arithmetic expressions (e.g., 8 − 6 and 6 / 3), and asked them to determine which of the two expressions was greater or if they were equal. Participants were given 25 seconds to complete as many comparisons as possible, with a new question presented immediately after each response. Visual feedback for correct and incorrect answers was provided. Participants were scored on the number of correct responses.

The VMT presents participants with a grid of light and dark squares for two seconds, after which the participant is asked to recreate the pattern on a blank grid (Della Sala et al., 1999). This test was repeated four times. During the first two trials 4 dark squares were randomly placed on a 3 x 3 grid, and during the second two trials 7 dark squares were randomly placed on a 4 x 4 grid. There was no time limit on this test, and audio and visual feedback were provided for correct and incorrect answers. Participants were scored on the total number of correct squares out of 22 possible correct answers.

## 2.5. Data collection

This study was made possible by the use of synchronization tools to collect and collate multiple data streams, specifically the Lab Streaming Layer (LSL) software package (Kothe, 2014). LSL incorporated data from multiple physiological sensors, participant task responses, and video screen recordings, as well as the insertion of event markers and participant responses. This video data was used to calculate the test response scores as well as to create a complete list of events to correlate with the physiological data.

EEG activity was measured using a 63-channel actiCHamp (Brain Products GmbH). Fifty-seven electrodes were devoted to scalp EEG recordings, an additional 4 electrodes to EOG data (two placed laterally 1 cm outside the eyes, and two vertically from the right eye), and the final 2 electrodes for EKG measurement (**Figure 1A**). The BrainVision Recorder software package was used for all electrophysiological data-recording, and EEGLAB software, including custom scripts in MATLAB (Mathworks, Natick, MA) was used for data-cleaning and analysis. The current analysis only involves the 57 scalp electrodes.

Participants interacted with the virtual environment by using an Xbox handheld controller. This features a small joystick that was used to toggle between multiple-choice answers, and buttons that were pressed to indicate the selection of the answer. Participants were able to look around the 3D space of the virtual classrooms by moving their heads, but they were not able to change their position within the virtual classroom. Head-accelerometer data was collected, but not analyzed or reported here.

**A. Four classroom design conditions**

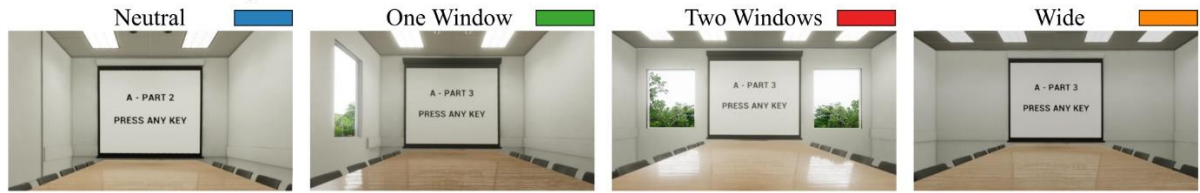

**B. Five cognitive tests preformed**

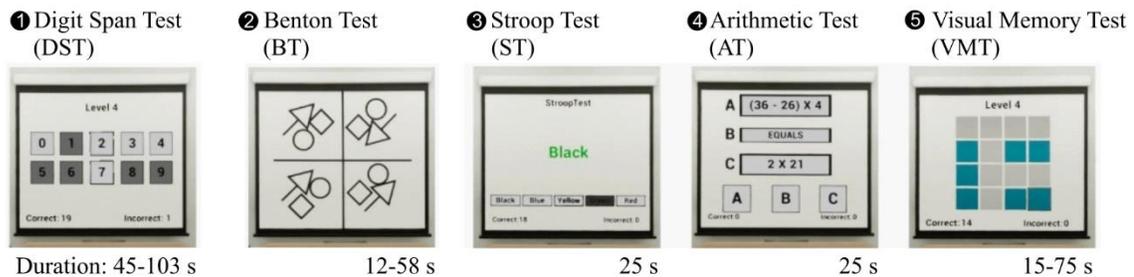

**C. Experiment timeline**

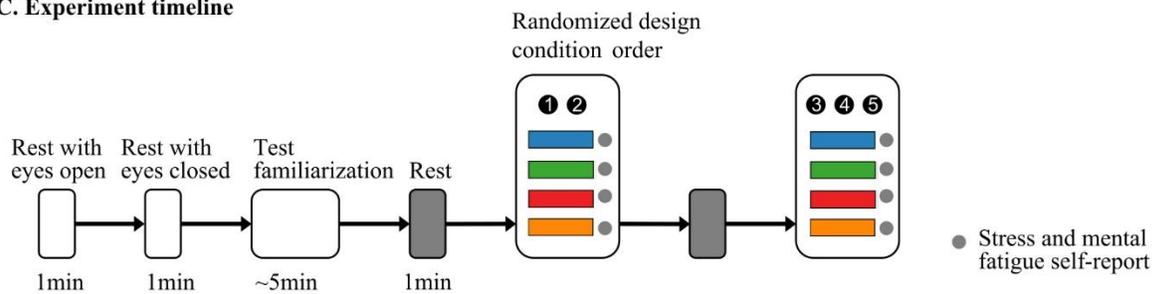

**Figure 2.** Overview of the study design. **A.** Four classroom design conditions were evaluated, and **B.** five cognitive tests were performed. **C.** The experiment's timeline included a familiarization period followed by exposure to each classroom environment in a random order.

## 2.6. Experimental procedure

Prior to beginning their individual session, each participant was provided with an introduction to the study and its goals, provided their written consent and completed a demographic questionnaire. The participant was then fitted with the EEG cap and the VR headset, and allowed to "play" with the headset using a non-test environment for a few minutes to become accustomed to the equipment. At the start of the formal experiment, baseline EEG readings were taken by asking the participant to focus on a spot on the wall for one minute with eyes open, and then to rest for one minute with eyes closed. The participant was then given a ten-minute period to become familiar with the cognitive tests to be performed and to practice the input methods for submitting test answers (**Figure 2C**).

The participant was then asked to complete the five cognitive tests in each different virtual classroom environment. The testing took place in two stages, with one minute of rest in between. In the first stage, participants completed the DST and then the BT in each of the four different classroom settings. While the order of the tasks was constant, the classroom variations appeared in a randomized sequence. After each classroom design, participants were asked to rate their mental fatigue and stress level on a scale from 1–10 (1="not at all" and 10="extremely"). After completing the first round of tasks, the participants were given a 60-second break with the headset on, during which time they were encouraged to relax while looking at a virtual beach scene.

In the second stage, participants were again asked to complete cognitive tasks in the different classrooms, this time including the Stroop test, the AT, and the VMT. Similar to the first stage, the order of the tests was constant while the order of the room conditions was randomized, and after each classroom the participants were asked to rate their mental fatigue and stress level using the same 1–10 scale. Commonly measured environmental indicators, such as heat, noise, and air quality, were held constant throughout the experiment, thus allowing for a controlled investigation of the effects of window placements and room dimensions.

### 2.7. Analysis of test performance data

Response accuracy, defined as the total number of correct responses divided by the total number of submitted responses, was utilized as a metric to evaluate the participants' performance on all cognitive tests. Additional test performance metrics were also measured: total number of correct responses, for the two tests that were time-limited at 25 s (the ST and AT), and total test duration, for those tests that were not time-limited (DST, BT, and VMT). For these latter three tests, the researchers also computed participants' Inverse Efficiency Scores following Townsend and Ashby (Townsend & Ashby, 1978). This is calculated as $IES = RT/(1-PE)$, where in our case RT is the overall task duration in one environment (not reaction-time), and PE is the proportion of error (1 – PercentCorrect). This is an appropriate measure to use when the proportion of error is low (Bruyer & Brysbaert, 2011; Vandierendonck, 2017), as was the case in the current experiment.

The distributions of the test performance metrics between the four classroom design conditions were compared using a Kruskal-Wallis test at a significance level of $p < 0.05$ and $p < 0.0025$ (Bonferroni-correction for $p < 0.05 / 20$ comparisons).

### 2.8. EEG data processing

The EEG data were analyzed using the EEGLAB software package (Delorme & Makeig, 2004). Raw .xdf data files were imported at their original sampling rate of 500 Hz. The data were band-pass filtered between 0.5 Hz and 50 Hz using the EEGLAB function pop_eegfiltnew (FIR, Hamming windowed, transition bandwidth 0.5 Hz).

Artifact subspace reconstruction (ASR) (Chang et al., 2019; C. A. E. Kothe & Jung, 2016) was carried out through its EEGLAB implementation in the function clean_artifacts. This function was used to remove channels that showed no signal activity (flat line threshold: 10 s), noisy signals (noisy line threshold: 4 std), or a poor correlation with adjacent channels (correlation threshold: 0.75). ASR was also used to remove short-time high amplitude artifacts in the continuous data, with a cut-off threshold of 20 standard deviations for identification of corrupted subspaces. Removed channels were then reintroduced to the data using spherical interpolation from neighboring channels. The data were common-average referenced over all of the scalp channels.

The data were then run through an independent component analysis (ICA) using the extended Infomax algorithm (Onton et al., 2006). Artifactual components representing eye-movements, head movements, and other gross non-brain components were removed automatically using the ICLabel EEGLAB plugin, with a label probability threshold of 0.90 for muscle, eye, heart, and channel-noise related artifacts (Pion-Tonachini et al., 2017, 2019). This automated pre-processing method was verified through visual inspection.

*2.9. Analysis of EEG features between design conditions*

The researchers examined EEG features in these preprocessed datasets during the first 25 s for each of the five cognitive tests, within each of the four environmental conditions (20 classes in total). We used various spectral (frequency-band power) and sensor-based connectivity EEG features to train a classifier to distinguish the neutral classroom condition from each of the other design variations. Thus, the three design condition comparisons were: (a) Neutral vs One-Window, (b) Neutral vs Two-Windows, and (c) Neutral vs Wide. These three comparisons were carried out separately for each of the five cognitive tests.

We chose a machine-learning classification approach to address the question of how different classroom environments affect cognitive performance for two reasons. First, we were agnostic to the potential cognitive impacts of these architectural design features, and machine-learning classifier methods can reveal statistically rigorous group differences even when a priori hypotheses are not clear. Second, given the possibility of finding no significant differences in task-performance across the classroom designs, a machine-learning approach was believed to be appropriate to help identify neural responses to the environments that might not correlate with behavioral performance in our current experimental paradigm.

### 2.9.1. EEG feature extraction

Two types of EEG features were analyzed at the sensor level: frequency band-power and partial directed coherence (PDC). These features were calculated over a 4-second sliding window with a 2-second overlap. The EEG features were taken as the relative change from baseline (1 min rest with eyes open) from each participant. The resulting features were each standardized by subtracting the mean and dividing by the standard deviation among all participants' data.

Frequency band-power features provide information about neural synchronization or desynchronization observed at scalp locations (Pfurtscheller & Lopes Da Silva, 1999; Sharma et al. 2017b). This results in an indication of how strongly a particular frequency range contributes to an EEG time-series signal. The Multitaper power spectral density estimate function (Thomson, 1982) in Matlab (pmtm) was used to calculate the frequency-band power, with a time-half-bandwidth product of 4. The band power features were calculated for all 57 electrodes for five typical EEG frequency bands: Delta (1–4 Hz), Theta (4–8 Hz), Alpha (8–12 Hz), Beta (12–30 Hz), and Gamma (30–40 Hz). This method produced a total of 285 frequency band-power features.

PDC provides a metric for functional interactions observed between electrodes (Cruz-Garza et al., 2020). It is based on the concept of partial coherence (Baccalá & Sameshima, 2001), a technique that quantifies the relationship between two signals while avoiding volume conduction. PDC measures directional (i.e. causal) influences between two EEG signals, which determines if there is transfer of information from one electrode to another. PDC was estimated using a multivariate autoregressive model (MVAR) using 28 electrodes distributed across the scalp, to reduce the number of comparisons to be made and still evaluating connectivity across the scalp. We used an MVAR model order of 15 (30 ms), which was obtained by using the ARFIT algorithm (Schneider and Neumaier, 2001) and evaluating the SBC criterion, which is least affected by noise (Porcaro et al., 2009). The observed PDC estimates were evaluated bidirectionally for all pairs of electrodes in the five frequency bands, yielding a total of 3920 PDC features.

### 2.9.2. EEG feature selection

The discriminant power of each EEG feature was estimated using the non-parametric Kruskal-Wallis test, which does not assume the data are normally distributed. The features across all study participants were ranked based on the $p$-values obtained from the Kruskal-Wallis tests for each two-class design condition comparison. The features that yielded a $p$-value smaller than $2.4 \times 10^{-6}$ (Bonferroni correction of $p < 0.01$ for 4205 features) were selected for visualization and analysis.

### 2.9.3. Machine learning classification

Support vector machines (SVMs) and linear discriminant analyses are the most common classifiers used in EEG-based brain-computer interfaces (Lotte et al., 2018). They are often used as baseline comparisons for the performance of emergent classifiers (Nakagome et al., 2020). The machine learning classifier used for the EEG data in the current experiment was the kernel support vector machine (k-SVM), with a polynomial kernel of degree 2 and box-constraint set to 1.

We defined 20 classes to be analyzed (based on five cognitive tests, each of which was conducted in four different classroom variations). The goal of the machine learning process was to evaluate if it could reliably differentiate a person's brain activity between two design settings; and if this differentiation was dependent on the cognitive test being performed (Hypothesis 1). To achieve class balance (Cruz-Garza et al., 2014, Hernandez et al., 2014), we randomly selected 40 data samples (where one data sample is a 4-second window of EEG data) from each participant for each of these 20 classes.

All comparisons were done in a pair-wise fashion (2-class problems), comparing the data samples from one of the three altered design conditions against data from the "neutral" classroom. The comparisons were done independently for each cognitive test. For each of the two-class comparisons, the most discriminant features from the EEG data were ranked, then sequentially added to the classification model and tested for predictive performance (Cruz-Garza et al., 2014). Along with the behavioral performance results (e.g. participant accuracy and reaction time, etc.), these EEG machine-learning analyses allowed the researchers to determine if the alterations in each classroom design had any significant effect on the neural correlates of participants' cognitive processes (Hypotheses 2, 3, and 4).

### 2.9.4. Training data set and testing data set

The machine learning procedure was conducted so that the identified EEG features could be tested for predictive power in an unseen participant. Thus, one of the participants was selected to be excluded from the training set. The data from the rest of the participants (n=22) was used as the training set to create the k-SVM classification model, using 5-fold cross-validation.

The process was iterated 23 times, excluding each participant in turn from the training set, and then evaluating the model based on its ability to predict the excluded individual's data. This approach provides strong verification of the model's predictive power. In each iteration, the training set contained 880 samples per class (22 participants), and the test set contained 40 samples per class (1 participant).

### 2.9.5. Classification model evaluation

The distribution of the classification accuracy for the testing set was used as the main metric for evaluating the machine learning classification performance. The distribution of the classification accuracy was compared to classification models and test data obtained from class-label scrambled EEG data (Agashe et al., 2015), within each comparison test. Statistical significance was evaluated between the classification accuracy in the test set and in the scrambled data with a Kruskal-Wallis test at a significance threshold of $p < 0.003$ and $p < 0.0006$ ($p < 0.05$ / 15 comparisons, and $p < 0.01$ / 15 comparisons, respectively), which applied the Bonferroni correction for multiple comparisons. The full machine learning process is summarized in **Figure 3.** The ranked EEG features were incrementally used to create classification models and test them to evaluate classification accuracy, using the top 400 best-ranked features.

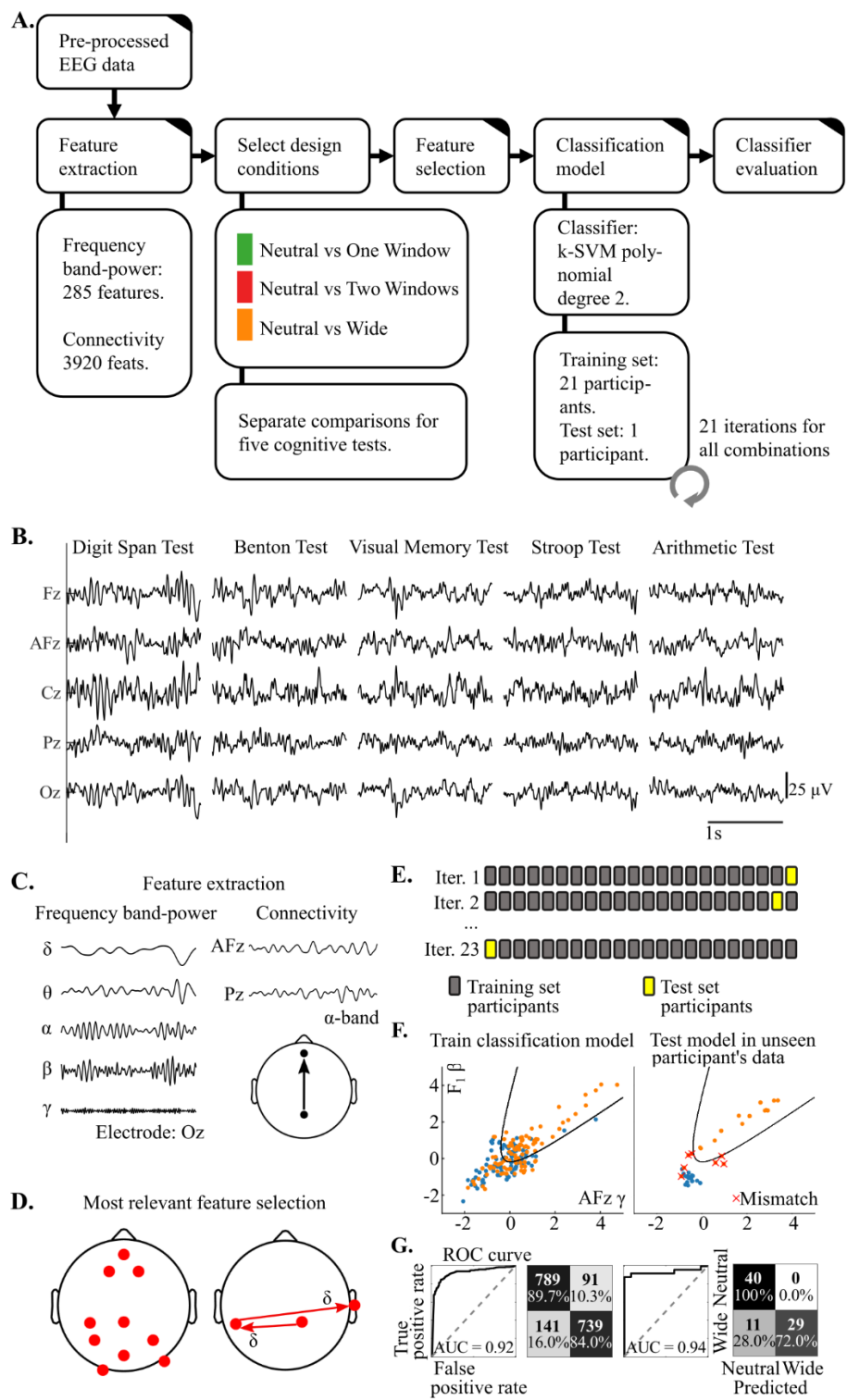

**Figure 3.** EEG data analysis and machine learning classification implementation, with sample data from one participant as it goes through the classification pipeline. **A.** EEG feature classification flowchart. The black marks on the top right corner indicate that a visual example of the process is shown in sections **B-G**. **B.** EEG data from one participant, in the Wide classroom design condition, for 2 s in all cognitive tests. **C.** Feature extraction: frequency band-power for each channel; and connectivity using PDC. **C.** The 14

most discriminant features to classify between the Neutral vs Wide design conditions. The classifier model shown corresponds to the VMT. **D.** Machine learning data distribution in the Training and Testing sets for 23 iterations. In each iteration, the data from an unseen participant was used for the Testing set. **E.** Machine learning example for one iteration. A classification model (k-SVM polynomial degree 2) was built from the Training set, and evaluated in the Testing set. The receiver operating characteristic curve (ROC), classification matrix (with number of samples and percentage representations), and classification accuracy are used to evaluate the classification performance.

# 3. RESULTS

## *3.1. Cognitive test performance results*

No overall significant effects were found for the different classroom conditions in relation to the participants' test response accuracy, the number of responses, completion time, or inverse efficiency scores, based on a Kruskal-Wallis test at a significance level of $p < 0.05$ and $p < 0.01$ (**Figure 4**). The response accuracy showed a strong skewness toward perfect scores for the BT, the VMT, the ST, and the AT. These results suggest that the participants were not challenged by these tasks, had achieved, via practice, mastery of them, and/or were very focused on completing these tasks successfully. The DST showed a broader distribution for the response accuracy metric, which may have been influenced by the feedback conditions and/or the nature of the test: if a participant selected a wrong number in the pattern, then frequently the following numbers selected would also be incorrect.

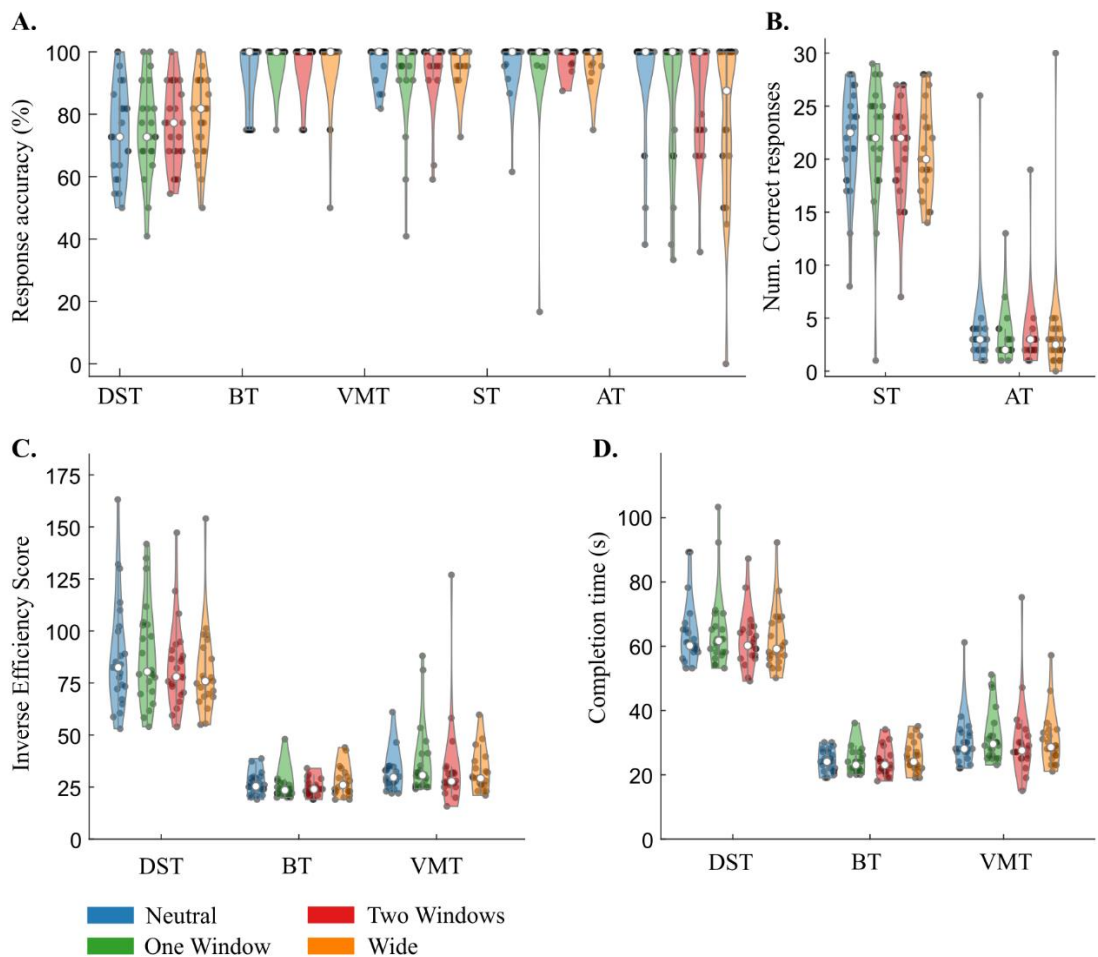

**Figure 4.** Cognitive test performance for all participants across the different classroom design conditions. No significant differences were found among the different classroom designs in regard to: **A.** response accuracy, **B.** total number of correct responses (for time-limited tests), **C.** Inverse Efficiency Scores, and **D.** completion time.

*3.2. EEG data-analysis results*

The machine learning method used frequency band-power and connectivity features of the EEG data to attempt to classify the different classroom design conditions. The resulting model was evaluated based on its ability to predict the EEG patterns of any participant that was excluded from the training set, in comparison against a control prediction based on scrambled data. The test-set classification accuracy distributions for each comparison are shown in **Figure 5**.

The classification results show that the different classroom design conditions were associated with consistent changes in the participants' EEG data, and that a classification model could be built with the predictive power to identify which design condition a new participant is experiencing. These results support Hypothesis 1.

The predictive power of the EEG features to identify the classroom design condition was specific to the cognitive test performed. When the participants were engaged in the VMT, the classifier could identify any of the three design conditions compared to the Neutral classroom with statistically significant success ($p < 0.003$). Specifically, the test-set median classification accuracies for the three design conditions evaluated were: 55.0% Neutral vs One-Window, 57.5% Neutral vs One-Window, 61.3% Neutral vs One-Window. The classifier also distinguished, significantly higher than chance ($p < 0.003$), the design conditions evaluated for the Benton Test in the Neutral vs One-Window comparison, and in the Neutral vs Two-Windows comparison; with test-set median classification accuracies of 61.3% and 56.3%, respectively.

The Digit Span Test also yielded EEG features that differed between the Neutral vs One-Window comparison, and in the Neutral vs Two-Windows comparison; with test-set median classification accuracies of 53.8% and 51.3%, respectively. These results partially support Hypothesis 2, 3, and 4, as the changes in EEG frequency band-power and connectivity associated with the design alterations were found only during the cognitive tests that required short-term memory encoding: the Digit Span Test, the Benton Test, and the Visual Memory Test. The ranked classification models were evaluated incrementally using 1 to 400 best-ranked features in each comparison. At 180 features, we observed the most comparisons reaching classification accuracies that were statistically significantly higher than chance levels (**Supplementary Materials Figure 1**).

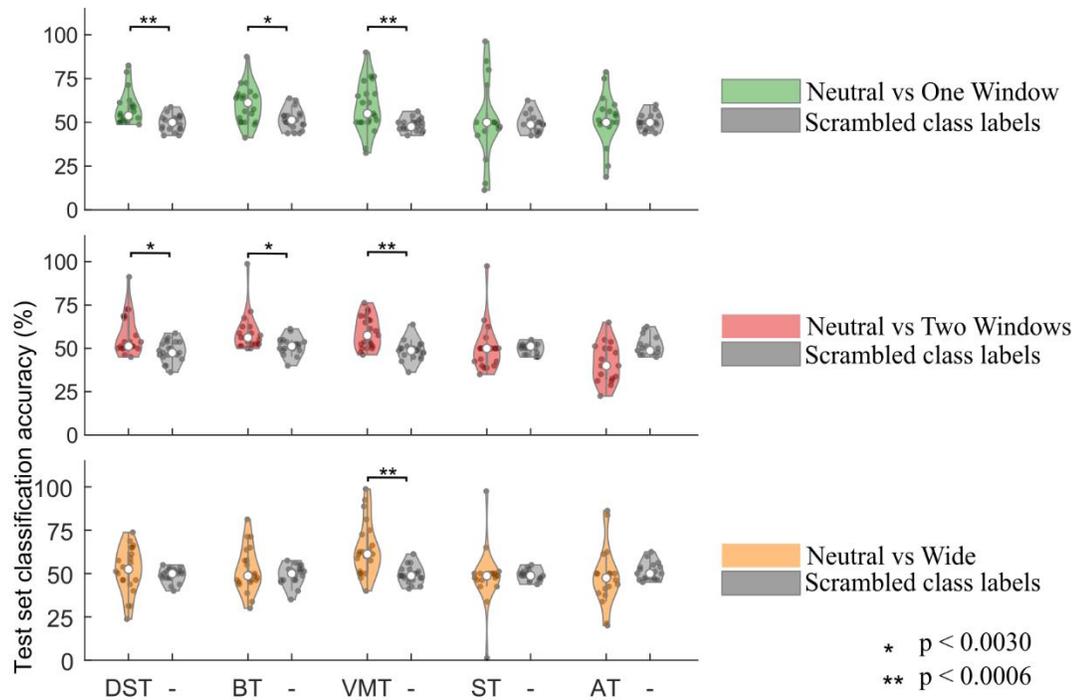

**Figure 5.** EEG classification accuracy results. A k-SVM (polynomial degree 2) classification model was created from the training data of 22 participants; and tested on unseen data from 1 participant- for 22 iterations. Each dot indicates the classification accuracy in one of the test sets based on the classification model obtained from its corresponding training set. To assess the validity of our results, we also calculated chance levels by scrambling the EEG data class labels in each 2-class comparison. The distribution for the scrambled class labels data is shown in gray. EEG classification accuracies are above the chance levels in most of the classes, and are significantly higher than the chance levels in 5 of the classes, supporting the validity of our method. Statistical significance was calculated using a pair-wise Kruskal-Wallis test.

The 180 most relevant features supporting the k-SVM classifiers where the classification models proved robust across participants are shown in **Figure 6**. This indicates the intra-participant changes in EEG as a percentage change from baseline "rest with eyes open", in response to the different classroom environments. Only those features that achieved a statistical significance threshold of $p < 2.4 \times 10^{-6}$ are displayed. The distribution of frequency band-power features showed significant differences in centro-parietal and frontal regions when they were compared between the Neutral design and the three alternative design conditions: One Window, Two Windows, and Wide classroom. Connectivity analysis reinforces this finding by showing that there are changes in the transfer of information from centro-parietal to frontal electrodes in the different classroom design conditions. Frequency band-power features were more discriminative between design conditions than the PDC connectivity features evaluated here. For the Neutral vs. One-Window comparison, the EEG band-power feature changes were characterized by an increase in relative alpha and beta power in the centro-parietal and frontal regions of the scalp. In contrast, the Neutral vs. Two Windows, and the Neutral vs Wide, both showed a decrease in relative alpha and beta power in these same regions. The VMT was the only cognitive test for which the occipital electrodes had the most discriminative power between the different design conditions (**Figure 6**).

The features that drive this classification performance are found in centro-parietal and frontal electrode locations (**Figure 6**). Additionally, connectivity analysis reinforces that there are changes in the transfer of information from centro-parietal to frontal electrodes. These features are found mostly in the theta and alpha frequency bands (4-12 Hz). The features that reach the statistical significance level, and their distributions, are different between the design conditions vs the neutral condition, are predominantly frequency band-power features. Connectivity features, measured using PDC, reach the threshold at a lower rate.

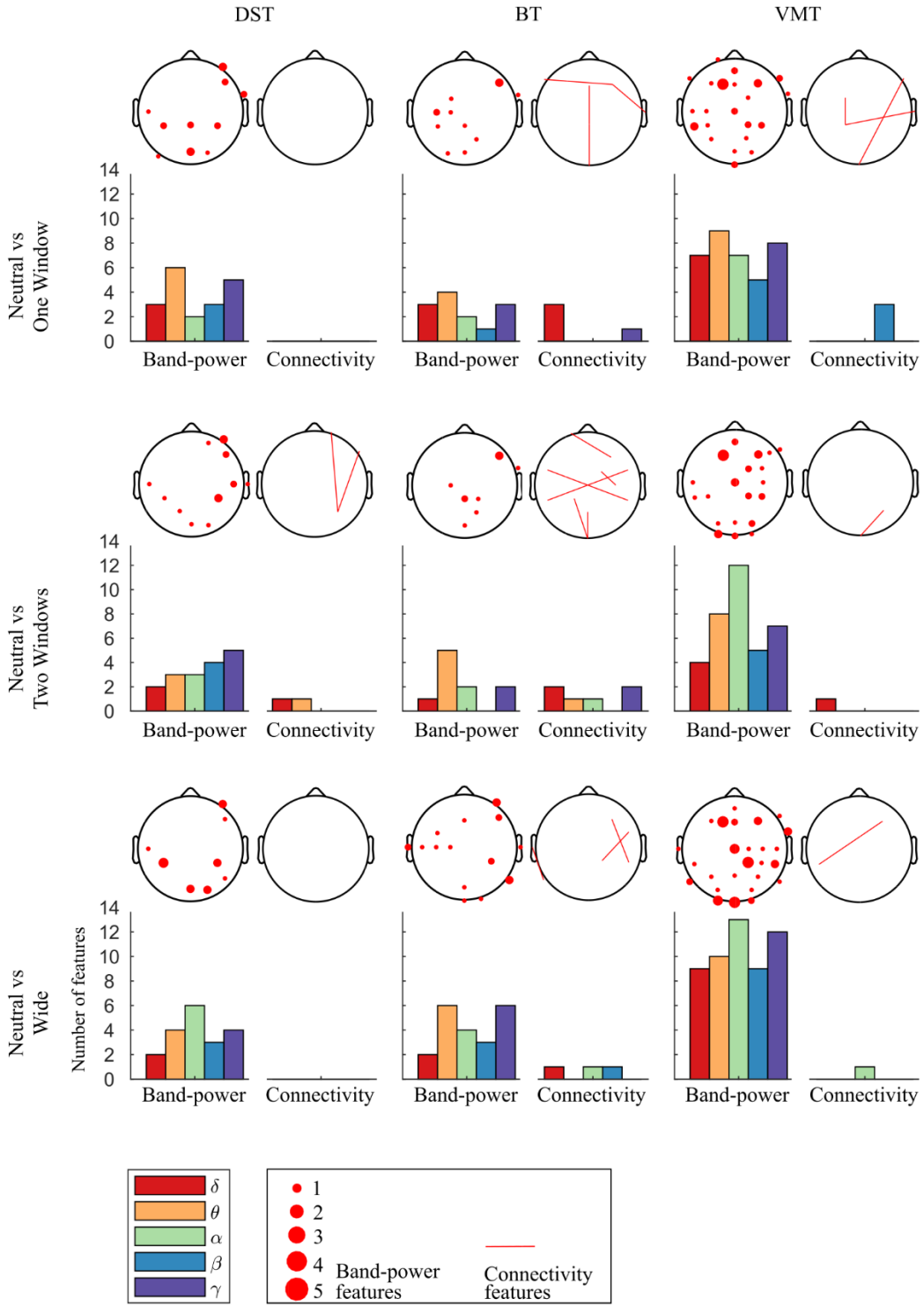

**Figure 6.** EEG statistically significant feature distribution. The number of features statistically significant features ($p < 2.4 \times 10^{-6}$ ) corresponding to frequency band-power and connectivity are displayed as bar graphs, for all five frequency bands analyzed. The scalp maps show the spatial distributions of the features , and the number of times an electrode reached the significance threshold.

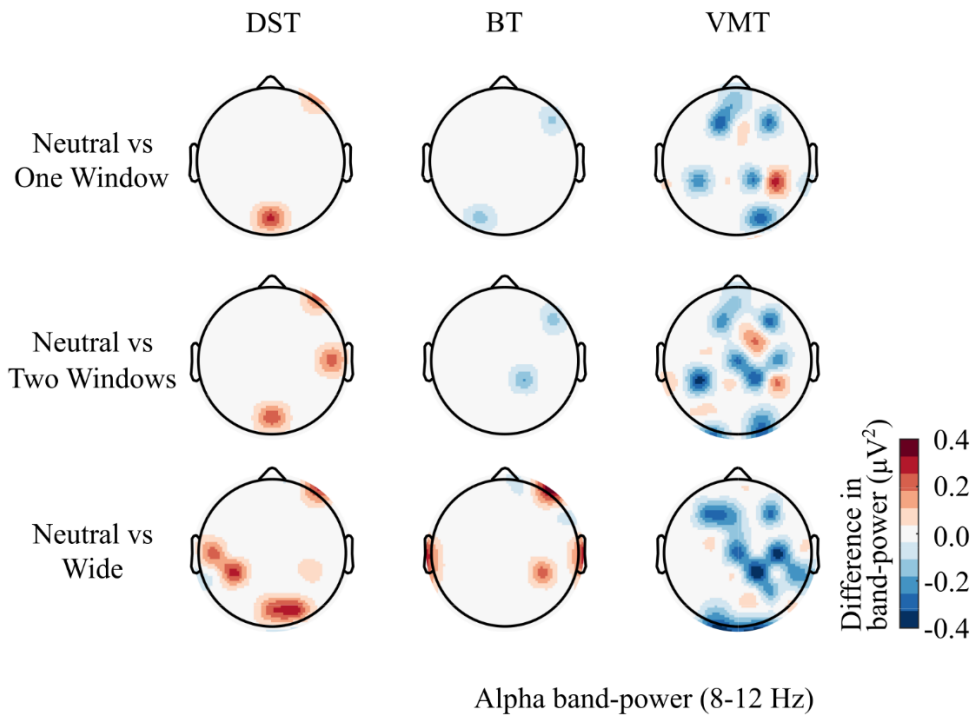

Alpha band-power (8-12 Hz)

**Figure 7.** EEG alpha band-power feature visualization. The topographic plots display the alpha band features that reached a statistical significance level of $p < 2.4 \times 10^{-6}$ ($p < 0.01$ Bonferroni-corrected for 4205 features). These plots show the difference between the design conditions compared with the Neutral classroom design; evaluated in three cognitive tests: DST, BT, and VMT.

The frequency band-power features were further visualized by plotting the difference in the features' distribution between the design conditions and the neutral classroom in specific frequency bands. In **Figure 7**, we display the difference in the distribution median for the alpha band. Only those comparisons that yielded robust classification performance in our inter-participant machine learning approach are displayed. The pattern of relevant features, synchronization and desynchronization in the alpha band, was different for each cognitive test evaluated; with an overall consistent pattern within the cognitive tests: the DST shows significant features in bilateral parietal and right frontal regions of the scalp; the BT shows the most significant features in centro-parietal regions and right-frontal regions; while the VMT shows significant features across a network of bilateral occipital, central, and medial frontal regions of the scalp. These results suggest that the assessment of an architectural design element in an indoor environment is dependent on the task performed by the participants.

For the DST, there was alpha-band synchronization within the parietal area, and among right-frontal electrodes, identified by an increase in band-power in the Design condition compared with the Neutral classroom (**Figure 7**). In the BT a similar pattern was found as for the DST, but parietal and right-frontal locations showed desynchronization in the alpha band in the design condition evaluated. The VMT was found to yield the largest number of features that reached the significance threshold. These features are distributed across the scalp, with an overall desynchronization in alpha power with the largest decrease in power in centro-parietal and mid-frontal electrodes. This overall desynchronization is coupled with localized right-parietal and right-frontal alpha-band synchronization.

## 4. DISCUSSION AND CONCLUSIONS

This study aimed to address logistical challenges that have limited the robust empirical study of indoor environment design and in particular classroom design. Using an immersive VR environment allowed for precise adjustments in isolated architectural variables, and was combined with an EEG-based approach to collect objective neural data from participants as they completed cognitive tests in these varied environments. A post-hoc machine-learning approach helped identify the EEG signals that best differentiated classroom architectural design conditions.

The cognitive test performance results showed no significant differences in accuracy or response times, response time, or related measures resulting from the architectural design features studied (window placement and classroom width). It is notable that this outcome is not commensurate with previous work that has found, for example, views of nature to be associated with improved working memory and decreased mental fatigue (Bratman et al., 2015; Gladwell et al., 2012; Li & Sullivan, 2016). One possible explanation for these contradictory results may be the length of exposure. Kaplan's (1995) Attention Restoration Theory and Ulrich and colleagues' (1991) Stress Reduction Theory both propose that the beneficial effects of nature enable involve the attenuation of certain cognitive, affective and physiological functions — e.g. directed attention (Kaplan), or stress response of the sympathetic nervous system (Ulrich and colleagues) – resulting from time spent viewing nature. In the current study, participants remained in each virtual classroom for less than two minutes and were constantly under mild stress while completing tests in quick succession. Thus, there may not have been enough contemplative time for the window views to have a restorative effect. It is also possible that the virtual environment may have omitted some of the holistic features responsible for cognitive restoration (such as robust complexity in the environment, a sense of physical/tactile relationship, and the conscious awareness of connections to nature), though additional research would be needed to verify if such factors are of primary significance. Another possible explanation for the lack of significant differences in the cognitive test results might be related to the fact that participants skewed strongly toward perfect scores (**Figure 4A**), which may result from their advanced cognitive ability as part of an academic. It is also potentially an indication of the participants' consistent engagement with the cognitive tasks, and/or or due to the practice effects of having substantial time to master each task before this evaluation.

The study found consistent and significant changes in EEG features during exposure to the architectural design variables, more prominently for cognitive tasks that required short-term memory encoding. This is a promising result, as it indicates a pathway toward the more robust scientific study of the effects of architectural design on cognitive processes. The fact that the classifier could successfully distinguish environments based on brain activity, *despite* no significant behavioral differences, in a sense strengthens this conclusion. Had the different classroom environments been associated with dramatically different response accuracies or reaction times, the classifier results could be interpreted as simply a reflection of participants' task performance. Instead, the classifier results can be interpreted as representing subtle cognitive, neural and/or psychological effects of the different architectural designs themselves. The leave-one-participant-out strategy is a robust method of cross-validating a classifier (Peng et al., 2019). The leave-one-out cross-validation strategy in the test set (unseen participant's data) provides a robust estimation for the generalization potential of the k-SVM classifier. This method produces a slightly different training dataset and classification boundary (**Figure 3F**) in each iteration of the training and test sets.

It is important to reflect on why the classroom design features which seemed to induce differentiated neural differences only did so in tasks that required maintaining presented stimuli in working memory (Benton, Visual Memory, and Digit Span; **Figure 5**). In contrast to the Stroop Test and the Arithmetic Test, which rely on executive control and arithmetic ability, respectively, the three working memory tasks (Benton, Visual Memory, and Digit Span) involve holding recently presented information in memory, in accordance with the encoding-specificity hypothesis. This likely involved storing some aspects of the context surrounding the task-relevant information (Staudigl & Hanslmayr, 2019), potentially strengthening the mental representations of the different classroom designs. It would follow that if the mental representation of the surrounding environment was stronger, a classifier could more effectively distinguish mental activity from short-term memory tasks occurring in those different environments. Additionally, it may not necessarily be just the mental representation of the task context that is strengthened. Other downstream cognitive processes may be subtly affected by each environment to the extent that does not dramatically affect the accuracy and response-time results, but which a sufficiently robust EEG classifier can detect.

Research on the encoding-specificity hypothesis contributes to this interpretation (Grissmann et al., 2020; Vogelsang et al., 2018; Zhou et al., 2017). The encoding specificity hypothesis proposes that memory recall is greater when the recall conditions or environment are similar to the original conditions or environment during memory encoding. A prominent related explanation to account for the ample evidence supporting the encoding specificity hypothesis is the neural reinstatement hypothesis, which proposes that the reoccurrence of neural activity during recall that was happening in the encoding stage should enhance and support memory recall (Crowley et al., 2019). This suggests that neural activity associated with a context and environment may be stronger when a memory task is involved than when a non-memory task is involved. Furthermore, the strongest encoding-specific effects are expected when the memory task and the context are in the same sensory modality (Staudigl & Hanslmayr, 2019). This supports our observation that the EEG classifier seemed to find more significantly classifiable neural features for the Benton Test and Visual Memory Test, than for the auditory Digit Span Test. This increased number of informative features in the VMT and BT, vs the DST, was observed across all three classroom comparisons (**Figure 7** and **Supplementary Materials Figure 1**), although further statistical analyses and experiments are needed to replicate and verify this observation. An open question remains, however, regarding whether the encoding-specificity hypothesis would account for classifier performance in "real-world" environments, in which design features are not carefully controlled. Regarding the general effect of each design on neural activity across cognitive tasks, further experiments and analyses are needed. Such experiments should use longer sessions in each design condition, and should measure baseline resting activity within each classroom environment, before and/or after doing tasks.

The specific features of the recorded EEG activity – the different frequency band-power values and transfer of information between different scalp electrodes – that were found in the present study to be most informative for the environment-classifier, converge with other theories about the role of the environment in task-specific cognitive functioning, such as the Parietal-Frontal Integration Theory (Vakhtin et al., 2014). Other source-localization methods, or studies using neuro-imaging modalities with a higher spatial resolution (e.g. fMRI), are needed in order to make strong conclusions about the role of specific anatomical and functional regions in cognitive performance differences caused by different environments. However, our machine-learning classifier results suggest that the Dorsal Attention Network is implicated, specifical connectivity between dorso-medial pre-frontal cortex and superior parietal areas (**Figure 6**; Spreng et al., 2013), including alpha power differences in expected visual and motor-planning networks (**Figure 7;** Zumer et al., 2014). However, the lack of behavioral differences between classrooms does suggest that the most predictive neural features did not involve performance-enhancing or task-performance-supporting neural activity. Nevertheless, it is possible that the practice effects – the fact participants had already achieved high proficiency in these tasks due to previous exposure – dominate the performance results. In that case, perhaps the different classrooms could lead to performance differences in the right conditions (such as prolonged exposure, or using less familiar or repetitive tasks), which in turn might change or enhance the classifiable different neural responses.

An additional set of questions to answer regards the specific effects of each classroom environment. First, we must consider that it is possible the unique demands of each task dominate the results, obscuring the cognitive effects of the different classroom designs. For example, among the tasks, the classifier found the most significant scalp regions when distinguishing the Visual Memory Test from the control classroom. We postulate (as discussed above) that this is explained in part by the encoding specificity hypothesis. Nevertheless, the classifier seemed to require unique EEG features for each experimental contrast, suggesting each classroom design had a unique influence on participants' neural dynamics. For example, a wider classroom seemed to modulate alpha power in left posterior tempero-parietal cortex compared to the control classroom in all three contrasts (wide > control, in the Digit Span Test; but wide < control in the two visual memory tests); this cortical area contributed no significant distinguishing alpha band features in the One-window and Two-window contrasts. Another example of a room-specific feature of interest is the seemingly bilateral distribution of occipital, central, and frontal cortex features (both spectral and connectivity features) that distinguished the Two-window condition, across cognitive tasks (**Figure 6**). The occipital bilaterality, as well as the parietal cross-hemispheric connectivity patterns, might be the result of the symmetrical distribution of the two windows across the visual field, but further work is needed to replicate and explain the bilateral distribution of relevant spectral features across other cortical areas.

The novel experimental/analytical approach and the results of this project may be of interest to researchers in multiple fields. Designers are likely to be excited to consider the wealth of useful findings that will emerge as this neural approach to evaluating user responses continues to evolve. The analysis conducted here, showing rigorous evidence that brain activity responses can be classified in association with specific architectural design elements, points toward a promising future for the use of physiological data in evidence-based design. This approach may someday be used to evaluate the impact of important environmental design decisions prior to physical construction, by collecting the physiological as well as conscious reactions of the intended user population. In the field of human neuroscience, the results of the current study expand on previous work showing that sensor-level feature extraction combined with machine learning can meaningfully distinguish environmental and cognitive variables, ultimately leading toward the implementation of more mobile, low-density EEG headsets to take the technology into real-world settings (Cruz-Garza et al., 2020; Ravindran et al., 2019; Rounds et al., 2020).

*4.1. Limitations*

The number of participants in this study was relatively small, which can increase the likelihood of a type II error (false negative findings). Furthermore, a large majority of the participants were male university students, with a relatively high degree of computer experience. A recent study on the effectiveness of VR learning as compared to hands-on learning reported that individuals who identified as male, and those having prior experience with computer games, performed better in the VR learning environment compared to their female and non-gamer counterparts, even when controlling for level of VR experience (Madden et al., 2020). This raises the possibility that responses to architectural designs in VR environments could be mediated by sex and/or gender, driven by prior VR experience. Future studies in the area of virtual classroom design can benefit from using larger and more diverse participant pools, with a special emphasis on considering gender and previous gaming experience as possible mediating variables. These factors should also be considered in ongoing work to confirm correspondences between real-world and VR neural patterns.

The cognitive tests that were used in this experiment are well studied and have been shown to have high reliability and validity in laboratory studies. However, the implications of such performance measures for long-term learning outcomes is not straightforward. Future studies may focus on context-relevant tasks that are more directly related to realistic learning experiences. The exposure duration for each environmental condition was very short in our study (1-2 minutes), which may have limited any restorative effect of the window views. Additionally, the "ceiling effect", in which many participants scored nearly perfectly on many tasks, limits some of the conclusions that can be drawn about the influence of these different environments.

*4.2. Future directions*

According to *The State of Technology in Education Report* (Promethean, 2020), virtual reality is now seen as the third most important growth area for technology in education, after cloud computing and online assessment, and its adoption will likely increase dramatically as a result of the COVID-19 pandemic. How should these new learning spaces be designed? Empirical tests of controlled changes in virtual environments could lead to the optimization of design elements to better support distance learning. Future studies in this area can examine even more granular interventions (for example by examining many small changes in the width, depth, and height of classrooms) to optimize such parameters for the general population, or even for specific demographic groups or individuals. It may turn out that different kinds of educational tasks and different student learning styles may be supported best by different classroom design features. In the virtual context where the environment can be quickly and readily adjusted, it may be possible for students to effortlessly switch between environments that optimize their performance on different tasks. It is even technologically feasible for each student to be immersed in a customized virtual learning space suited to his or her needs, while still interacting with teachers and peers in their own customized environments. The optimization of these virtual environments will lead to findings that can later be translated and tested in the physical world.

# References

Adam, G. E., Carter III, R., Cheuvront, S. N., Merullo, D. J., Castellani, J. W., Lieberman, H. R., & Sawka, M. N. (2008). Hydration effects on cognitive performance during military tasks in temperate and cold environments. *Physiology & Behavior*, *93*(4–5), 748–756. https://doi.org/10.1016/j.physbeh.2007.11.028

Agashe, H. A., Paek, A. Y., Zhang, Y., & Contreras-Vidal, J. L. (2015). Global cortical activity predicts shape of hand during grasping. *Frontiers in Neuroscience*, *9*, 121. https://doi.org/10.3389/fnins.2015.00121

Baccalá, L. A., & Sameshima, K. (2001). Partial directed coherence: A new concept in neural structure determination. *Biological Cybernetics*, *84*(6), 463-474. https://doi.org/10.1007/PL00007990

Bagot, K. L., Allen, F. C. L., & Toukhsati, S. (2015). Perceived restorativeness of children's school playground environments: Nature, playground features and play period experiences. *Journal of Environmental Psychology, 41*, 1-9. https://doi.org/10.1016/j.jenvp.2014.11.005

Banaei, M., Hatami, J., Yazdanfar, A., & Gramann, K. (2017). Walking through architectural spaces: The impact of interior forms on human brain dynamics. *Frontiers in Human Neuroscience*, *11*, 477. https://doi.org/10.3389/fnhum.2017.00477

Bandelow, S., Maughan, R., Shirreffs, S., Ozgünen, K., Kurdak, S., Ersöz, G., Binnet, M., & Dvorak, J. (2010). The effects of exercise, heat, cooling and rehydration strategies on cognitive function in football players. *Scandinavian Journal of Medicine & Science in Sports*, *20*, 148–160. https://doi.org/10.1111/j.1600-0838.2010.01220.x

Barrett, P., Davies, F., Zhang, Y., & Barrett, L. (2015). The impact of classroom design on pupils' learning: Final results of a holistic, multi-level analysis. *Building and Environment, 89,* 118-133 https://doi.org/10.1016/j.buildenv.2015.02.013

Benfield, J. A., Rainbolt, G. N., Bell, P. A., & Donovan, G. H. (2015). Classrooms With Nature Views: Evidence of Differing Student Perceptions and Behaviors. *Environment and Behavior*, *47*(2), 140-157, https://doi.org/10.1177/0013916513499583

Benton, A. L. (1945). A visual retention test for clinical use. *Archives of Neurology And Psychiatry*, *54*(3), 212–216. https://doi.org/10.1001/archneurpsyc.1945.02300090051008

Bratman, G. N., Daily, G. C., Levy, B. J., & Gross, J. J. (2015). The benefits of nature experience: Improved affect and cognition. *Landscape and Urban Planning, 138*, 41-50. https://doi.org/10.1016/j.landurbplan.2015.02.005

Bruyer, R., & Brysbaert, M. (2011). Combining speed and accuracy in cognitive psychology: Is the inverse efficiency score (IES) a better dependent variable than the mean reaction time (RT) and the percentage of errors (PE)? *Psychologica Belgica, 51*(1), 5-13. https://doi.org/10.5334/pb-51-1-5

Cama, R. (2009). *Evidence-based healthcare design*. John Wiley & Sons.

Castellucci, H. I., Arezes, P. M., Molenbroek, J. F. M., de Bruin, R., & Viviani, C. (2017). The influence of school furniture on students' performance and physical responses: Results of a systematic review. *Ergonomics, 60*(1), 93-110. https://doi.org/10.1080/00140139.2016.1170889

Chang, C.-Y., Hsu, S.-H., Pion-Tonachini, L., & Jung, T.-P. (2019). Evaluation of artifact subspace reconstruction for automatic artifact components removal in multi-channel EEG recordings. *IEEE Transactions on Biomedical Engineering*, *67*(4), 1114–1121.

Chen, J. (2016). *Attention restoration benefits of integrating green space into office environment*. [Doctoral dissertation, University of Illinois at Urbana-Champaign]. http://hdl.handle.net/2142/93067

Choi, Y., Kim, M., & Chun, C. (2015). Measurement of occupants' stress based on electroencephalograms (EEG) in twelve combined environments. *Building and Environment, 88*, 65-72. https://doi.org/10.1016/j.buildenv.2014.10.003

Choi, Y., Kim, M., & Chun, C. (2019). Effect of temperature on attention ability based on electroencephalogram measurements. *Building and Environment, 147,* 299-304. https://doi.org/10.1016/j.buildenv.2018.10.020

Choo, H., Nasar, J. L., Nikrahei, B., & Walther, D. B. (2017). Neural codes of seeing architectural styles. *Scientific Reports, 7*(1), 1-8. https://doi.org/10.1038/srep40201

Coburn, A., Vartanian, O., & Chatterjee, A. (2017). Buildings, beauty, and the brain: A neuroscience of architectural experience. *Journal of Cognitive Neuroscience, 29*(9), 1521-1531. https://doi.org/10.1162/jocn_a_01146

Collado, S., & Corraliza, J. A. (2015). Children's Restorative Experiences and Self-Reported Environmental Behaviors. *Environment and Behavior, 47*(1), 38-56. https://doi.org/10.1177/0013916513492417

Contreras-Vidal, J. L., Robleto, D., & Cruz-Garza, J. G. (2019). Towards a Roadmap for Neuroaesthetics. In *Mobile Brain-Body Imaging and the Neuroscience of Art, Innovation and Creativity* (pp. 215–220). Springer.

Crowley, R., Bendor, D., & Javadi, A.-H. (2019). A review of neurobiological factors underlying the selective enhancement of memory at encoding, consolidation, and retrieval. *Progress in Neurobiology*, *179*, 101615. https://doi.org/10.1016/j.pneurobio.2019.04.004

Cruz-Garza, Jesus G, Brantley, J. A., Nakagome, S., Kontson, K., Robleto, D., & Contreras-Vidal, J. L. (2017). Mobile EEG Recordings in an Art Museum Setting. *IEEE Dataport*. https://doi.org/10.21227/H2TM00

Cruz-Garza, J. G., Hernandez, Z. R., Nepaul, S., Bradley, K. K., & Contreras-Vidal, J. L. (2014). Neural decoding of expressive human movement from scalp electroencephalography (EEG). *Frontiers in Human Neuroscience*, 8(188), 188. https://doi.org/10.3389/fnhum.2014.00188

Cruz-Garza, J. G., Sujatha Ravindran, A., Kopteva, A. E., Rivera Garza, C., & Contreras-Vidal, J. L. (2020). Characterization of the stages of creative writing with mobile EEG using Generalized Partial Directed Coherence. *Frontiers in Human Neuroscience*, *14*, 533. https://doi.org/10/ghnns7

Daisey, J. M., Angell, W. J., & Apte, M. G. (2003). Indoor air quality, ventilation and health symptoms in schools: An analysis of existing information. *Indoor Air, 13*(LBNL-48287). https://doi.org/10.1034/j.1600-0668.2003.00153.x

Della Sala, S., Gray, C., Baddeley, A., Allamano, N., & Wilson, L. (1999). Pattern span: A tool for unwelding visuo-spatial memory. *Neuropsychologia*, *37*(10), 1189–1199. https://doi.org/10.1016/S0028-3932(98)00159-6

Delorme, A., & Makeig, S. (2004). EEGLAB: An open source toolbox for analysis of single-trial EEG dynamics including independent component analysis. *Journal of Neuroscience Methods*, *134*(1), 9–21. https://doi.org/10.1016/j.jneumeth.2003.10.009

Demetriou, A., Kazi, S., Makris, N., & Spanoudis, G. (2020). Cognitive ability, cognitive self-awareness, and school performance: From childhood to adolescence. *Intelligence 79*, 101432. https://doi.org/10.1016/j.intell.2020.101432

Dikker, S., Wan, L., Davidesco, I., Kaggen, L., Oostrik, M., McClintock, J., Rowland, J., Michalareas, G., Van Bavel, J. J., Ding, M., & Poeppel, D. (2017). Brain-to-Brain Synchrony Tracks Real-World Dynamic Group Interactions in the Classroom. *Current Biology, 27*(9), 1375–1380. https://doi.org/10.1016/j.cub.2017.04.002

Edelstein, E. A., & Macagno, E. (2012). Form follows function: Bridging neuroscience and architecture. In *Springer Optimization and Its Applications*. https://doi.org/10.1007/978-1-4419-0745-5_3

Engelbrecht, K. (2003). The Impact of Color on Learning. *NeoCon Perkins & Will*.

Faber Taylor, A., & Kuo, F. E. (2009). Children With Attention Deficits Concentrate Better After Walk in the Park. *Journal of Attention Disorders*, *12*(5), 402–409. https://doi.org/10/btbppn

Gladwell, V. F., Brown, D. K., Barton, J. L., Tarvainen, M. P., Kuoppa, P., Pretty, J., Suddaby, J. M., & Sandercock, G. R. H. (2012). The effects of views of nature on autonomic control. *European Journal of Applied Physiology*, *112*(9), 3379–3386. https://doi.org/10/fx4t92

Glen I. Earthman. (2004). Prioritization of 31 Criteria for School Building Adequacy. *Manual*.

Grissmann, S., Spüler, M., Faller, J., Krumpe, T., Zander, T. O., Kelava, A., Scharinger, C., & Gerjets, P. (2020). Context Sensitivity of EEG-Based Workload Classification Under Different Affective Valence. *IEEE Transactions on Affective Computing*, *11*(2), 327–334. https://doi.org/10.1109/taffc.2017.2775616

Guan, H., Hu, S., Lu, M., He, M., Zhang, X., & Liu, G. (2020). Analysis of human electroencephalogram features in different indoor environments. *Building and Environment*, *186*, 107328. https://doi.org/10/ghm25n

Hamilton, D. K., & Watkins, D. H. (2008). *Evidence-based design for multiple building types*. John Wiley & Sons.

Harvey, E. J., & Kenyon, M. C. (2013). Classroom seating considerations for 21st century students and faculty. *Journal of Learning Spaces*, *2*(1).

Hernandez, Z. R. Z. R., Cruz-Garza, J., Tse, T., & Contreras-Vidal, J. L. J. L. (2014). Decoding of intentional actions from scalp electroencephalography (EEG) in freely-behaving infants. 2014 36th Annual International Conference of the IEEE Engineering in Medicine and Biology Society, EMBC 2014, 2014. https://doi.org/10.1109/EMBC.2014.6944034

Kalantari, S., Contreras-Vidal, J. L., Smith, J. S., Cruz-Garza, J., & Banner, P. (2018). Evaluating educational settings through biometric data and virtual response testing. *Recalibration on Imprecision and Infidelity - Proceedings of the 38th Annual Conference of the Association for Computer Aided Design in Architecture, ACADIA 2018*, 118–125.

Kalantari, S. (2019, July). A new method of human response testing to enhance the design process. In *Proceedings of the Design Society: International Conference on Engineering Design* (Vol. 1, No. 1, pp. 1883-1892). Cambridge University Press.

Kalantari, S., Rounds, J. D., Julia, K., Vidushi, T., & Cruz-Garza, J. G. (2020). Immersive virtual environments vs. Identical real-world environments: Comparing physiological responses during cognitive tests. *Scientific Reports*, *In press*.

Kalantari, S., & Neo, J. R. J. (2020). Virtual environments for design research: lessons learned from use of fully immersive virtual reality in interior design research. *Journal of Interior Design*, *45*(3), 27-42.

Kaplan, S. (1995). The restorative benefits of nature: Toward an integrative framework. *Journal of Environmental Psychology*, *15*(3), 169–182. https://doi.org/10/dxkgxh

Kidger, J., Araya, R., Donovan, J., & Gunnell, D. (2012). The effect of the school environment on the emotional health of adolescents: A systematic review. *Pediatrics*. https://doi.org/10.1542/peds.2011-2248

Kim, H., Hong, T., Kim, J., & Yeom, S. (2020). A psychophysiological effect of indoor thermal condition on college students' learning performance through EEG measurement. *Building and Environment, 184*, 107223. https://doi.org/10.1016/j.buildenv.2020.107223

Kjellgren, A., & Buhrkall, H. (2010). A comparison of the restorative effect of a natural environment with that of a simulated natural environment. *Journal of Environmental Psychology*, *30*(4), 464–472. https://doi.org/10.1016/j.jenvp.2010.01.011

Klatte, M., Hellbrück, J., Seidel, J., & Leistner, P. (2010). Effects of Classroom Acoustics on Performance and Well-Being in Elementary School Children: A Field Study. *Environment and Behavior*, *42*(5), 659–692. https://doi.org/10/bw656q

Ko, W. H., Schiavon, S., Zhang, H., Graham, L. T., Brager, G., Mauss, I., & Lin, Y.-W. (2020). The impact of a view from a window on thermal comfort, emotion, and cognitive performance. *Building and Environment*, *175*, 106779. https://doi.org/10/ghnn6m

Kothe, C. (2014). Lab streaming layer (LSL). *Https://Github. Com/Sccn/Labstreaminglayer. Accessed on October*.

Kothe, C. A. E., & Jung, T. P. (2016). *U.S. Patent Application No. 14/895,440*.

Küller, R., Mikellides, B., & Janssens, J. (2009). Color, arousal, and performance-A comparison of three experiments. *Color Research & Application*, *34*(2), 141–152. https://doi.org/10/b2n422

Li, D., & Sullivan, W. C. (2016). Impact of views to school landscapes on recovery from stress and mental fatigue. *Landscape and Urban Planning*, *148*, 149–158. https://doi.org/10/f8gh4d

Lotte, F., Bougrain, L., Cichocki, A., Clerc, M., Congedo, M., Rakotomamonjy, A., & Yger, F. (2018). A review of classification algorithms for EEG-based brain–computer interfaces: A 10 year update. *Journal of Neural Engineering*, *15*(3), 031005. https://doi.org/10/gf48q9

Lu, M., Hu, S., Mao, Z., Liang, P., Xin, S., & Guan, H. (2020). Research on work efficiency and light comfort based on EEG evaluation method. *Building and Environment*, *183*, 107122. https://doi.org/10/ghjp3z

Madden, J., Pandita, S., Schuldt, J. P., Kim, B., S. Won, A., & Holmes, N. G. (2020). Ready student one: Exploring the predictors of student learning in virtual reality. *PLOS ONE*, *15*(3), e0229788. https://doi.org/10/ghnn63

Mäkinen, T. M., Palinkas, L. A., Reeves, D. L., Pääkkönen, T., Rintamäki, H., Leppäluoto, J., & Hassi, J. (2006). Effect of repeated exposures to cold on cognitive performance in humans. *Physiology & Behavior*, *87*(1), 166–176. https://doi.org/10.1016/j.physbeh.2005.09.015

Makransky, G., Andreasen, N. K., Baceviciute, S., & Mayer, R. E. (2020). Immersive virtual reality increases liking but not learning with a science simulation and generative learning strategies promote learning in immersive virtual reality. *Journal of Educational Psychology*. https://doi.org/10/ggrwxg

Martin, K., McLeod, E., Périard, J., Rattray, B., Keegan, R., & Pyne, D. B. (2019). The Impact of Environmental Stress on Cognitive Performance: A Systematic Review. *Human Factors*, *61*(8), 1205–1246. https://doi.org/10.1177/0018720819839817

Matsuoka, R. H. (2010). Student performance and high school landscapes: Examining the links. *Landscape and Urban Planning*, *97*(4), 273–282. https://doi.org/10/b7rff9

Maxwell, L. E., & Schechtman, S. L. (2012). The Role of Objective and Perceived School Building Quality in Student Academic Outcomes and Self-Perception. *Children, Youth and Environments*, *22*(1), 23. https://doi.org/10/ghnn68

Mendell, M. J., & Heath, G. A. (2005). Do indoor pollutants and thermal conditions in schools influence student performance? A critical review of the literature. *Indoor Air*, *15*(1), 27–52. https://doi.org/10/b54r9q

Moore, G. T., Lane, C. G., Hill, A. B., Cohen, U., McGinty, T., Jules, F. A., ... & Work, L. L. (1996). *Recommendations for child care centers*. Center for Architecture and Urban Planning Research, University of Wisconsin--Milwaukee.

Morley, J., Beauchamp, G., Suyama, J., Guyette, F. X., Reis, S. E., Callaway, C. W., & Hostler, D. (2012). Cognitive function following treadmill exercise in thermal protective clothing. *European Journal of Applied Physiology*, *112*(5), 1733–1740. https://doi.org/10.1007/s00421-011-2144-4

Nakagome, S., Craik, A., Ravindran, A., He, Y., Cruz-Garza, J. G., & Contreras-Vidal, J. L. (2020). Deep learning methods for EEG neural classification. In N. V. Thakor (Ed.), *Handbook of Neuroengineering* (1st ed., p. In press). Springer.

Nigg, J. T., Quamma, J. P., Greenberg, M. T., & Kusche, C. A. (1999). A two-year longitudinal study of neuropsychological and cognitive performance in relation to behavioral problems and competencies in elementary school children. *Journal of abnormal child psychology*, *27*(1), 51-63.

Olds, A. R. (1989). Psychological and physiological harmony in child care center design. *Children's Environments Quarterly, 6*(4), 8–16.

Onton, J., Westerfield, M., Townsend, J., & Makeig, S. (2006). Imaging human EEG dynamics using independent component analysis. *Neuroscience and Biobehavioral Reviews*, *30*(6), 808–822. https://doi.org/10.1016/j.neubiorev.2006.06.007

Palanica, A., Lyons, A., Cooper, M., Lee, A., & Fossat, Y. (2019). A comparison of nature and urban environments on creative thinking across different levels of reality. *Journal of Environmental Psychology*, *63*, 44–51. https://doi.org/10.1016/j.jenvp.2019.04.006

Parker, S. M., Erin, J. R., Pryor, R. R., Khorana, P., Suyama, J., Guyette, F. X., Reis, S. E., & Hostler, D. (2013). The Effect of Prolonged Light Intensity Exercise in the Heat on Executive Function. *Wilderness & Environmental Medicine*, *24*(3), 203–210. https://doi.org/10.1016/j.wem.2013.01.010

Peng, H., Xia, C., Wang, Z., Zhu, J., Zhang, X., Sun, S., Li, J., Huo, X., & Li, X. (2019). Multivariate Pattern Analysis of EEG-Based Functional Connectivity: A Study on the Identification of Depression. IEEE Access, 7, 92630–92641. https://doi.org/10/ghnszt

Pion-Tonachini, L., Kreutz-Delgado, K., & Makeig, S. (2019). ICLabel: An automated electroencephalographic independent component classifier, dataset, and website. *NeuroImage*, *198*, 181–197. https://doi.org/10/gf22bz

Pion-Tonachini, L., Makeig, S., & Kreutz-Delgado, K. (2017). Crowd labeling latent Dirichlet allocation. *Knowledge and Information Systems*, *53*(3), 749–765. https://doi.org/10/gcgfmn

Porcaro, C., Zappasodi, F., Rossini, P. M., & Tecchio, F. (2009). Choice of multivariate autoregressive model order affecting real network functional connectivity estimate. Clinical Neurophysiology, 120(2), 436–448. https://doi.org/10/cs8trp

Promethean. (2020). *The State of Technology in Education Report*. https://resourced.prometheanworld.com/technology-education-industry-report

Radcliffe, D., Wilson, W., Powell, D., & Tibbetts, B. (Eds.). (2008). Learning spaces in higher education. Positive outcomes by design. In *Proceedings of the next generation learning spaces 2008 colloquium.* St Lucia, QLD: The University of Queensland.

Sujatha Ravindran, A., Mobiny, A., Cruz-Garza, J. G., Paek, A., Kopteva, A., & Contreras Vidal, J. L. (2019). Assaying neural activity of children during video game play in public spaces: A deep learning approach. *Journal of Neural Engineering*, *16*(3), 036028. https://doi.org/10/ghgkjn

Read, M. A., Sugawara, A. I., & Brandt, J. A. (1999). Impact of Space and Color in the Physical Environment on Preschool Children's Cooperative Behavior. *Environment and Behavior*, *31*(3), 413–428. https://doi.org/10/csdzdb

Roe, J. J., Aspinall, P. A., Mavros, P., & Coyne, R. (2013). Engaging the brain: The impact of natural versus urban scenes using novel EEG methods in an experimental setting. *Environmental Sciences*, *1*, 93–104. https://doi.org/10/ghnn7h

Rounds, J. D., Cruz-Garza, J. G., & Kalantari, S., (2020). Using Posterior EEG Theta Band to Assess the Effects of Architectural Designs on Landmark Recognition in an Urban Setting. *Frontiers in Human Neuroscience*, *14*, 537.

Salthouse, T. A. (2005). Relations Between Cognitive Abilities and Measures of Executive Functioning. *Neuropsychology*, *19*(4), 532–545. https://doi.org/10/fvvrc3

Schneider, T., & Neumaier, A. (2001). Algorithm 808: ARfit—a matlab package for the estimation of parameters and eigenmodes of multivariate autoregressive models. *ACM Transactions on Mathematical Software*, *27*(1), 58–65. https://doi.org/10/bbs2xh

Sharma, G., Kaushal, Y., Chandra, S., Singh, V., Mittal, A. P., & Dutt, V. (2017). Influence of Landmarks on Wayfinding and Brain Connectivity in Immersive Virtual Reality Environment. *Frontiers in Psychology*, *8*, 1220. https://doi.org/10/gbqh82

Shin, Y.-B., Woo, S.-H., Kim, D.-H., Kim, J., Kim, J.-J., & Park, J. Y. (2015). The effect on emotions and brain activity by the direct/indirect lighting in the residential environment. *Neuroscience Letters*, *584*, 28–32. https://doi.org/10/f6wqrt

Sivan, A. B. (1992). *Benton visual retention test*. Psychological Corporation San Antonio, TX.

Spitznagel, M. B., Updegraff, J., Pierce, K., Walter, K. H., Collinsworth, T., Glickman, E., & Gunstad, J. (2009). Cognitive Function During Acute Cold Exposure With or Without Sleep Deprivation Lasting 53 Hours. *Aviation, Space, and Environmental Medicine*, *80*(8), 703–708. https://doi.org/10.3357/asem.2507.2009

Spreng, R. N., Sepulcre, J., Turner, G. R., Stevens, W. D., & Schacter, D. L. (2013). Intrinsic Architecture Underlying the Relations among the Default, Dorsal Attention, and Frontoparietal Control

Networks of the Human Brain. *Journal of Cognitive Neuroscience*, *25*(1), 74–86. https://doi.org/10.1162/jocn_a_00281

Staudigl, T., & Hanslmayr, S. (2019). Reactivation of neural patterns during memory reinstatement supports encoding specificity. *Cognitive Neuroscience*, *10*(4), 175–185. https://doi.org/10.1080/17588928.2019.1621825

Stroop, J. R. (1935). Studies of interference in serial verbal reactions. *Journal of Experimental Psychology*, *18*(6), 643–662. https://doi.org/10/b77m95

Sundstrom, E. (1975). An experimental study of crowding: Effects of room size, intrusion, and goal blocking on nonverbal behavior, self-disclosure, and self-reported stress. *Journal of Personality and Social Psychology*, *32*(4), 645–654. https://doi.org/10/c67mps

Taylor, L., Fitch, N., Castle, P., Watkins, S., Aldous, J., Sculthorpe, N., Midgely, A., Brewer, J., & Mauger, A. (2014). Exposure to hot and cold environmental conditions does not affect the decision making ability of soccer referees following an intermittent sprint protocol. *Frontiers in Physiology*, *5*. https://doi.org/10/ghm43k

Tennessen, C. M., & Cimprich, B. (1995). Views to nature: Effects on attention. *Journal of Environmental Psychology*, *15*(1), 77–85. https://doi.org/10/fgzxbm

Thomson, D. J. (1982). Spectrum estimation and harmonic analysis. *Proceedings of the IEEE*, *70*(9), 1055–1096. https://doi.org/10/fcfqm4

Townsend, J. T., & Ashby, F. G. (1978). Methods of modeling capacity in simple processing systems. *Cognitive theory*, *3*, 199-139.

Ulrich, R. S. (2006). Essay: Evidence-based health-care architecture. *The Lancet*, *368*, S38–S39. https://doi.org/10/dgj5gt

Ulrich, R. S., Simons, R. F., Losito, B. D., Fiorito, E., Miles, M. A., & Zelson, M. (1991). Stress recovery during exposure to natural and urban environments. *Journal of Environmental Psychology*, *11*(3), 201–230. https://doi.org/10/csp47k

Vandierendonck, A. (2017). A comparison of methods to combine speed and accuracy measures of performance: A rejoinder on the binning procedure. *Behavior Research Methods*, *49*(2), 653–673. https://doi.org/10/f96whp

Vartanian, O., Navarrete, G., Chatterjee, A., Fich, L. B., Gonzalez-Mora, J. L., Leder, H., Modroño, C., Nadal, M., Rostrup, N., & Skov, M. (2015). Architectural design and the brain: Effects of ceiling height and perceived enclosure on beauty judgments and approach-avoidance decisions. *Journal of Environmental Psychology*, *41*, 10–18. https://doi.org/10.1016/j.jenvp.2014.11.006

Vartanian, O., Navarrete, G., Chatterjee, A., Fich, L. B., Leder, H., Modrono, C., Nadal, M., Rostrup, N., & Skov, M. (2013). Impact of contour on aesthetic judgments and approach-avoidance decisions in architecture. *Proceedings of the National Academy of Sciences*, *110*(Supplement_2), 10446–10453. https://doi.org/10/swv

Vecchiato, G., Tieri, G., Jelic, A., De Matteis, F., Maglione, A. G., & Babiloni, F. (2015). Electroencephalographic Correlates of Sensorimotor Integration and Embodiment during the Appreciation of Virtual Architectural Environments. *Frontiers in Psychology*, *6*. https://doi.org/10.3389/fpsyg.2015.01944

Vogelsang, D. A., Gruber, M., Bergström, Z. M., Ranganath, C., & Simons, J. S. (2018). Alpha Oscillations during Incidental Encoding Predict Subsequent Memory for New "Foil" Information. *Journal of Cognitive Neuroscience*, *30*(5), 667–679. https://doi.org/10.1162/jocn_a_01234

Wannarka, R., & Ruhl, K. (2008). Seating arrangements that promote positive academic and behavioural outcomes: A review of empirical research. *Support for Learning*, *23*(2), 89–93. https://doi.org/10/bsg596

Wargocki, P. (2019). Productivity and Health Effects of High Indoor Air Quality. In *Encyclopedia of Environmental Health* (pp. 382–388). Elsevier. https://doi.org/10.1016/B978-0-12-409548-9.01993-X

Wargocki, P., & Wyon, D. (2007). The Effects of Moderately Raised Classroom Temperatures and Classroom Ventilation Rate on the Performance of Schoolwork by Children (RP-1257). *HVAC&R Research*, *13*(2), 193–220. https://doi.org/10/dg9gb4

Wechsler, D. (1981). Manual for the Wechsler Adult Intelligence Scale—Revised. In *Psychological Corporation*.

Winchip, S. (1991). Factors Contributing to a Safe, Supportive and Desirable Housing Environment for Children. *Housing and Society*, *18*(3), 23–29. https://doi.org/10/ghnn8w

Winterbottom, M., & Wilkins, A. (2009). Lighting and discomfort in the classroom. *Journal of Environmental Psychology*, *29*(1), 63–75. https://doi.org/10/bbsnkh

Yin, J., Yuan, J., Arfaei, N., Catalano, P. J., Allen, J. G., & Spengler, J. D. (2020). Effects of biophilic indoor environment on stress and anxiety recovery: A between-subjects experiment in virtual reality. *Environment International*, *136*, 105427. https://doi.org/10/ggmddb

Yin, J., Zhu, S., MacNaughton, P., Allen, J. G., & Spengler, J. D. (2018). Physiological and cognitive performance of exposure to biophilic indoor environment. *Building and Environment*, *132*, 255–262. https://doi.org/10.1016/j.buildenv.2018.01.006

Zhang, F., de Dear, R., & Hancock, P. (2019). Effects of moderate thermal environments on cognitive performance: A multidisciplinary review. *Applied Energy*, *236*, 760–777. https://doi.org/10.1016/j.apenergy.2018.12.005

Zhou, H., Xiong, G.-J., Jing, L., Song, N.-N., Pu, D.-L., Tang, X., He, X.-B., Xu, F.-Q., Huang, J.-F., Li, L.-J., Richter-Levin, G., Mao, R.-R., Zhou, Q.-X., Ding, Y.-Q., & Xu, L. (2017). The interhemispheric CA1 circuit governs rapid generalisation but not fear memory. *Nature Communications*, *8*(1), 2190. https://doi.org/10.1038/s41467-017-02315-4

Zhu, M., Liu, W. & Wargocki, P. (2020) Changes in EEG signals during the cognitive activity at varying air temperature and relative humidity. *J Expo Sci Environ Epidemiol*, *30*, 285–298 https://doi.org/10.1038/s41370-019-0154-1

Zimring, C., Joseph, A., & Choudhary, R. (2004). The role of the physical environment in the hospital of the 21st century: A once-in-a-lifetime opportunity. *Concord, CA: The Center for Health Design*, *311*.

Zumer, J. M., Scheeringa, R., Schoffelen, J.-M., Norris, D. G., & Jensen, O. (2014). Occipital Alpha Activity during Stimulus Processing Gates the Information Flow to Object-Selective Cortex. *PLoS Biology*, *12*(10). https://doi.org/10.1371/journal.pbio.1001965

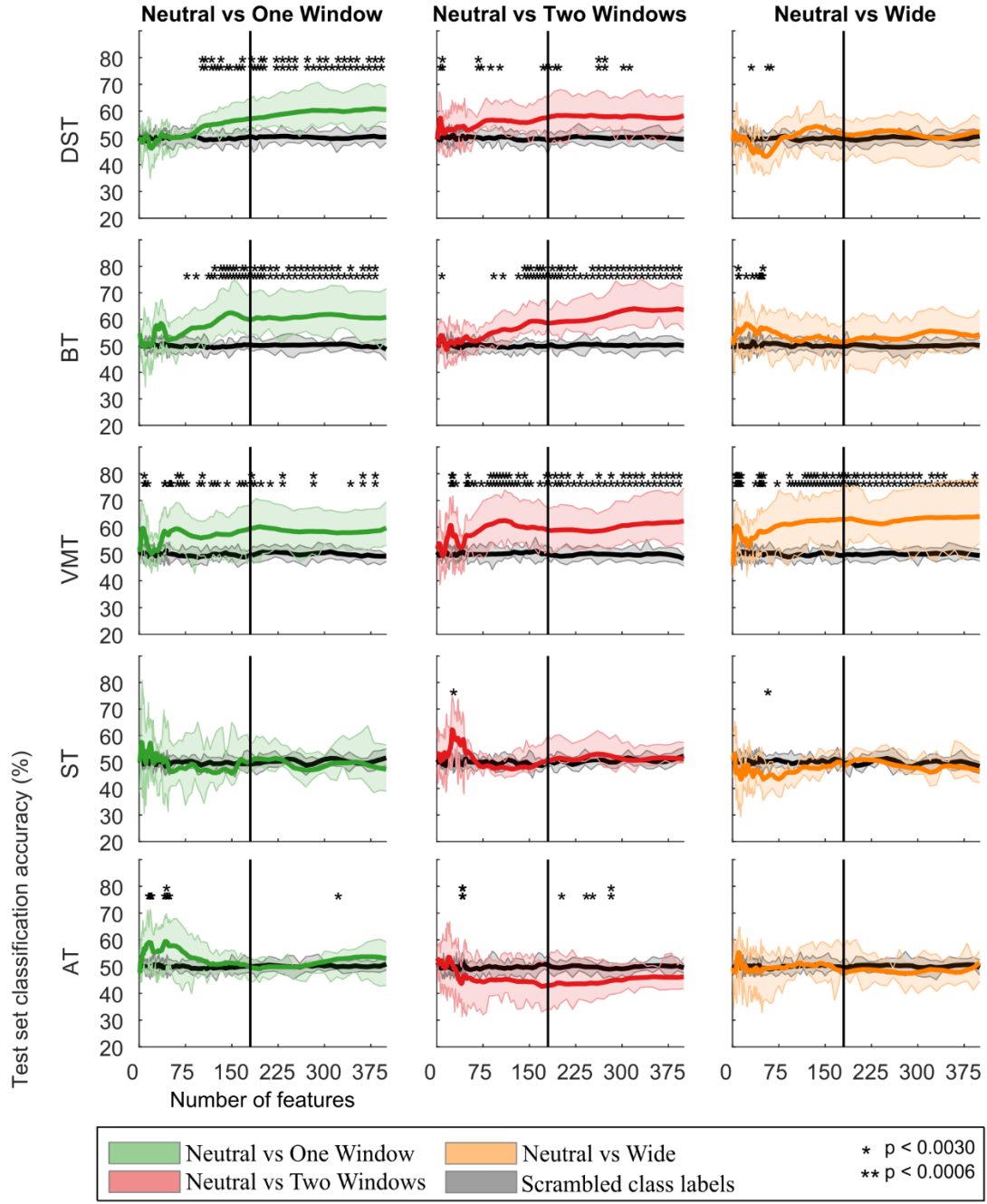

**Supplementary Materials Figure 1.** Classification accuracy between design conditions. To assess the validity of our results, we calculated chance levels by scrambling the time-window index that identifies the room the participants were in. * $p < 0.003$, ** $p < 0.0006$ ($p < 0.05$ and $p < 0.01$ Bonferroni-corrected for 15 comparisons) . The shaded areas indicate the interquartile range in classification accuracy. The vertical line is located at 180 features, for which most of the significant results were found among the comparisons.